\documentclass[10pt,twocolumn]{article} 

\usepackage{graphicx}
\usepackage{color}
\usepackage{hyperref}
\usepackage{soul}
\usepackage{amssymb}

\newcommand{\figu}[1]{\ref{#1}}
\newcommand{\equ}[1]{(\ref{#1})}

\begin{document}

\twocolumn[
\title{Innovation and nested preferential growth in chess playing behavior}
\maketitle

\author{\noindent J.I. Perotti$^{1,3}$, H.-H. Jo$^1$, A.L. Schaigorodsky$^2$ and O.V. Billoni$^2$\\
            \and
        {\em \small $^1$ Department of Biomedical Engineering and Computational Science, Aalto University School of Science - P.O. Box 12200, Finland\\
            \and
       $^2$ Facultad de Matem\'atica, Astronom\'ia y F\'isica, Universidad Nacional de C\'ordoba and Instituto de F\'isica Enrique Gaviola (IFEG-CONICET) - Ciudad Universitaria, 5000 C\'ordoba, Argentina}\\
        {\small $^3$ juanpool@gmail.com}
}


\abstract{
{\bf \small Complexity develops via the incorporation of innovative properties.
Chess is one of the most complex strategy games, where expert contenders exercise decision making by imitating old games or introducing innovations.
In this work, we study innovation in chess by analyzing how different move sequences are played at the population level. 
It is found that the probability of exploring a new or innovative move decreases as a power law with the frequency of the preceding move sequence. 
Chess players also exploit already known move sequences according to their frequencies, following a preferential growth mechanism. 
Furthermore, innovation in chess exhibits Heaps' law suggesting similarities with the process of vocabulary growth. 
We propose a robust generative mechanism based on nested Yule-Simon preferential growth processes that reproduces the empirical observations. 
These results, supporting the self-similar nature of innovations in chess  are important in the context of decision making in a competitive scenario, and extend the scope of relevant findings recently discovered regarding the emergence of Zipf's law in chess.}
}
]

\vspace*{0.5cm}

\section{Introduction}
In the last decades, the study of complex systems gained interest in the physics community, including studies on financial markets~\cite{potters1998financial}, bird flocks~\cite{toner1998flocks,bialek2012statistical}, the human brain~\cite{eguiluz2005scalefree}, cities~\cite{encarnacao2012fractal,bettencourt2013origins}, and elections~\cite{fortunato2007scaling} among other subjects.
Using methods borrowed from statistical physics, nonlinear dynamics, and network theory, researchers have improved the understanding of the structure, properties, and behavior of complex phenomena~\cite{amaral2004complex}.
However, how complex systems emerge and develop is still an important open question.
Complexity results from adaptation processes through which systems acquire innovative properties~\cite{wagner2011origins}.
Therefore, how and when innovations occur in complex phenomena constitutes an important subject to be understood.

An innovative event occurs when a new idea, product, lexicon, or gene is introduced and eventually exploited by the society or ecosystem. 
Novelties become innovations only when they bring utility or satisfaction to the members of the system in question. 
Innovation is necessary for the growth and development of organizations and economies~\cite{fagerberg2005oxford}, and it is believed to operate at the heart of natural evolution~\cite{wagner2011origins} and language evolution~\cite{bobda1994lexical,wildgen2004evolution}. 
An innovative behavior may enhance the chances of success in competitive scenarios~\cite{christensen1997innovator}. 
However, it requires the investment of limited resources in testing new possibilities instead of exploiting well known solutions. 
For this reason, it is important to understand how nature and society optimize between the exploration of new possibilities and the exploitation of old ones. 
In particular, this question is relevant in the context of human decision making when bounded rationality operates~\cite{simon1947administrative,arthur1994inductive}.

The game of chess~\cite{murray1986history} has been extensively studied in 
science~\cite{blasius2009zipf,sigman2010response,ribeiro2013move,schaigorodsky2014memory}, especially for decision making~\cite{blasius2009zipf,sigman2010response,de2011flexible}, and it can be used to study innovation phenomena. 
Chess is a competitive game with a game tree complexity estimated as $10^{120}$ different move sequences~\cite{shannon1950xxii}. 
Because of such huge number, all possible games cannot be fully searched in practice, thus it is necessary to continuously explore new possible moves. 
This implies that bounded rationality and innovation should operate. 
In addition, the quantitative characterization of innovation phenomena has been difficult mainly due to the lack of reliable datasets. 
This difficulty could be circumvented by the recent appearance of extensive records of chess games. 

In a recent work, Blasius and T\"onjes~\cite{blasius2009zipf} analyzed a dataset of chess games by mapping move sequences  or games to vertices in a game tree~\cite{shannon1950xxii}.
In particular, they studied how frequently each vertex in the tree is visited by a game, showing that the distribution of frequencies follows the Zipf's law~\cite{george1949zipf,newman2005power} over six orders of magnitude.
They explained this finding using a multiplicative process with fragmentation that can be treated analytically.
In their approach, each game develops following the branches of already explored parts of the game tree, and thus they focused on the dynamics within games.
However, the characterization of the growing dynamic of game tree in itself is still an open problem and its characterization is crucial for the understanding of the role of innovation in chess.

In this Letter, we investigate the evolution of the game tree to find the underlying processes which describe how innovation works in chess.
Questions such as when and where new vertices appear in the already explored game tree, or how the associated frequencies grows, are empirically addressed and answered.
These investigations have allowed us to propose a simple generative mechanism reproducing the observations of an extensive chess database.

\section{Results}
\begin{figure}[!t]
\includegraphics*[width=\linewidth]{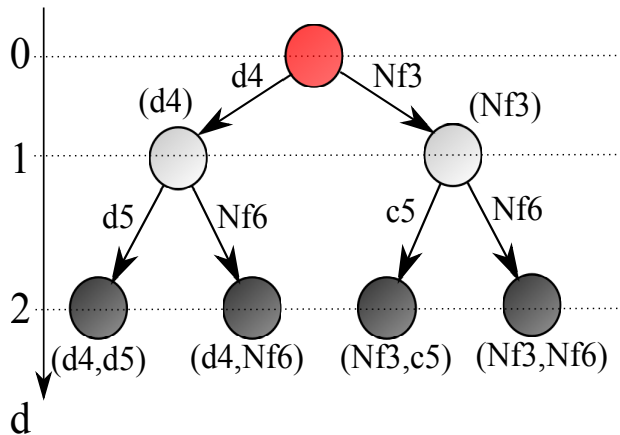}
\caption{\label{fig:1}
(Color online). Schematic representation of the chess game tree. The root (red vertex on top) represents the initial position on the board. The white (black) vertices correspond to positions after the white (black) player moved. Each directed edge corresponds to a move. After each move, the game depth $d$ increases by one. Each edge (vertex) is labeled by a move (a move sequence) using algebraic chess notation. Note that the same move, such as Nf6, may appear more than once in different edges.}
\end{figure}

We analyze a dataset including around 1.4 million chess games played between 1998 and 2007 in {\em ChessDB}~\cite{chessdb}. Each possible move sequence corresponds to one directed path in a game tree, starting from the root vertex or the initial position of the game. See fig.~\figu{fig:1}. The moves are represented by edges. Note that there is one-to-one correspondence between move sequences and vertices in the game tree. We grow an initially empty game tree by sequentially adding the chronologically sorted games in the dataset. The depth $d$ of a vertex is its distance from the root. We define that an innovative event occurs whenever a game generates a new branch in the existing tree. The $t$th game may generate a new branch at depth $d_b(t)$ and will end at depth $d_e(t)\geq d_b(t)$. Here $t$ plays a role of ordinal time or time in short. 
Note that each game may introduce only one new branch or none, where $d_b(t)$ is not defined for the latter case. 

\begin{figure}[!t]
\includegraphics*[width=\linewidth]{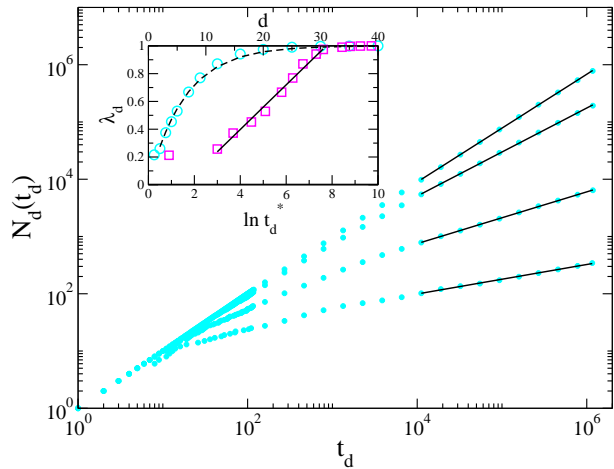}
\caption{\label{fig:2}
(Color online). Tree width $N_d$ versus the number of games $t_d$ that reached depth $d$ for the game tree in the chess dataset (cyan circles). We only show the cases of $d=2,4,9,16$ for clarity and the data is logarithmically binned for $t_d > 100$. Fitted results of the form $N_d = t_d^*(t/t_d^*)^{\lambda_d}$ for large values of $t_d$ are shown in black solid lines. {\em Inset:} The estimated values of $\lambda_d$ are plotted as a function of $d$ (cyan circles) and as a function of the estimated values of $\ln t_d^*$ (magenta squares). The black dashed line corresponds to the fit $\lambda_d = 1-B^d$ where $B=0.854\pm0.002$ ($R^2=0.998$), and the black solid line to $\lambda_d = a+(1/A) \ln t_d^*$ (see eq.~\ref{eq:crossover}) where $a=-0.24\pm0.04$ and $1/A=0.160\pm0.006$ ($R^2=0.99$).}
\end{figure}

For each innovative event occurring at depth $d_b(t)$, the tree width or number of vertices $N_d$ at depth $d$ evolves as $N_d(t)=N_d(t-1)+1$ if $d_b(t)<d\leq d_e(t)$ or $N_d(t)=N_d(t-1)$ otherwise. Let $t_d(t)$ be the number of games that reached at least depth $d$ after $t$ games have been played. From now on, the variable $t$ is omitted if not necessary. As shown in fig.~\figu{fig:2}, we find that
\begin{equation}
\label{eq:heaps}
N_d(t_d) \simeq \left\{\begin{array}{ll}
t_d & t_d \ll t_d^* \\
t_d^* (t_d/t_d^*)^{\lambda_d} & t_d \gg t_d^*,
\end{array}\right.
\end{equation}
where $t_d^*$ is a crossover value of $t_d$. The exponent $\lambda_d$, characterizing the innovation rate, saturates exponentially with $d$ as 
\begin{equation}
\label{eq:lambda_d}
\lambda_d = 1-B^d
\end{equation}
with $B\simeq 0.85$ (fig. \figu{fig:2}, Inset). The scaling $N_d \sim t_d^{\lambda_d}$ for $t_d \gg t_d^*$ corresponds to the Heaps' law commonly found in the vocabulary growth of literary corpora or languages~\cite{heaps1978information,serrano2009modeling,gerlach2013stochastic}. Moreover, it is found that the crossover $t_d^*$ exponentially grows with the Heaps' exponent as
\begin{equation}
\label{eq:crossover}
t_d^* \sim \exp(A\lambda_d).
\end{equation}
Double-scaling behavior in eq.~\equ{eq:heaps} was also found by Gerlach and Altmann~\cite{gerlach2013stochastic} in the context of vocabulary growth. They showed that this double-scaling is associated with a double-scaling in Zipf's law~\cite{van2005formal,zanette2005dynamics,cattuto2009collective,lu2010zipf,eliazar2011growth}. 
We confirm this double-scaling in the depth-dependent Zipf's laws in our dataset (not shown).


\begin{figure}[!t]
\includegraphics*[width=\linewidth]{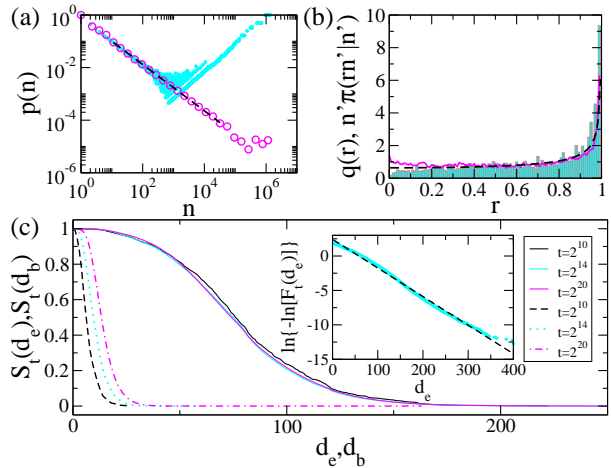}
\caption{\label{fig:3}
(Color online).
(a) The probability of generating a new branch, $p(n)$, as a function of the vertex frequency $n$ (cyan filled circles). 
The log-binned data (magenta circles) is fitted by $p(n) \sim n^{-\nu}$ (black dashed line) with $\nu=0.881\pm0.009$ ($R^2=0.9998$). 
(b) Empirical probability density function, $q(r)$, of the frequency ratio $r=n/n'$ (cyan histogram) and its analytical expression, $q(r)=2/[\pi\sqrt{1-r^2}]$ (black dashed line), compared with the rescaled form $n'\pi(rn'|n')$ of the conditional growth probability distribution, $\pi(n|n')$ (magenta continuous line).
(c) Fraction $S_t(d_e)$ ($S_t(d_b)$) of games with game lengths (branching depths) larger or equal to $d_e$ ($d_b$) after $t$ games have been played are plotted as solid lines (broken lines). 
Different colors indicate different values of $t$. 
The curves of $S_t(d_e)$ collapse into one, while the curves of $S_t(d_b)$ depend on $t$. 
{\em Inset:} Fit of a Gumbel distribution for maxima $F_t(d_e)=1-S_t(d_e)=\exp(-\exp(-(d_e-\mu_e)/\beta_e))$ (black dashed line) with $\mu_e=58.2\pm 0.2$ and $\beta_e=24.1\pm 0.1$ ($R^2=0.995$) for the largest $t$ case (cyan circles).}
\end{figure}

In order to understand the underlying mechanism of the evolution of the game tree, we characterize the innovation and exploitation processes. 
Let $n(t)$ be the frequency of occurrence of a given move sequence after $t$ games were played.
For the innovation process, the probability $p(n)$ that a game reaching a vertex with frequency $n$ generates a new branch turns out to be
\begin{equation}
\label{eq:prob_branch}
p(n) \simeq n^{-\nu},
\end{equation}
where $\nu \simeq 0.88$ (fig. \figu{fig:3}a). 
For the exploitation process, we study the conditional probability, $\pi(n|n')$, that a game follows an existing edge from a vertex with frequency $n'$ to one of its child vertices which has frequency $n$.
We measure $\pi(n|n')$ as the games traverse the edges of the growing tree.
For instance, fig.~\figu{fig:3}b (continuous magenta line) shows $\pi(n|n'=100)$ obtained from moves with depths $d\geq 5$ in order to assert an adequate statistics. 
Similar results are obtained for other values of $n'$ (not shown).
We found that $\pi(n|n')$ is an homogeneous function satisfying
\begin{equation}
\label{eq:pref_growth}
\pi(n|n') = \frac{1}{n'} q\left(\frac{n}{n'}\right),
\end{equation}
where $q(r)$ is the probability density function of frequency ratios, $r=n/n'$ (fig. \figu{fig:3}b, cyan histogram).
The probability density, $q(r)$, is measured along the growing process of the tree.
Moves ending in vertices with $n<100$ or $d < 5$ are discarded from the statistics in order to avoid discretization effects in the shape of $q(r)$.
The functional form $q(r)=2/[\pi\sqrt{1-r^2}]$ (fig. \figu{fig:3}b, black dashed line) was previously determined by Blasius and T\"onjes~\cite{blasius2009zipf} measuring the values of $r$ as the games traverse the already grown tree.
As $q(r)$ is an increasing function of $r$, branches with larger frequencies are played more frequently, and therefore it corresponds to a preferential growth process.
However, it is different from the typical case where the preferential growth/attachment probability grows linearly with the frequency~\cite{yule1925mathematical,simon1955class,price1976general,barabasi1999emergence,jeong2003measuring}.
We remark that the scaling behaviors of eqs.~\equ{eq:prob_branch} and~\equ{eq:pref_growth} have been measured over the whole tree, and therefore corresponds to the behavior of a depth independent process.
This, and the functional form of $\pi(n|n')$, evidences the self-similar nature of the evolution of the game tree.

Next, we consider the effect of the growing stage of the tree on the statistical properties of the game length $d_e$. 
The fraction $S_t(d_e)$ of games that did not end until length $d_e$ follows a Gumbel distribution for maxima~\cite{de2006extreme}, and it is independent of the number of games that have been played (fig.~\figu{fig:3}c, Inset).
On the other hand, the fraction $S_t(d_b)$ of games that did not branch until depth $d_b$ does depend on $t$ (fig. \figu{fig:3}c). 
Therefore, it turns out that game lengths are almost independent of the growing stage of the tree.

Now we introduce theoretical considerations in order to understand the mechanism that generates the game tree. 
We assume that games are infinite in length and therefore $t_d = t$ for all $d$. This assumption is justified by our previous observation about the statistical independence between the game lengths and the tree growth.
From now on, we consider the asymptotic behavior for large $t$. According to a depth-dependent mean field approach, the branching factor $K_d$ at depth $d$ satisfies
\begin{equation}
\label{eq:branching_factor}
K_d(t) = N_{d+1}(t)/N_d(t) \sim t^{B^d(1-B)},
\end{equation}
and the average frequency per vertex at depth $d$ is given by
\begin{equation}
\label{eq:average_frequency}
n_d = t/N_d(t) \sim t^{B^d}.
\end{equation}
Combining eqs.~\equ{eq:branching_factor} and \equ{eq:average_frequency} we obtain
\begin{equation}
\label{eq:branch_fact_vs_popu}
K_d \sim n_d^{1-B}.
\end{equation}
The branching factor at depth $d$ grows sub-linearly with the frequency $n_d$. 
The derivative of the branching factor with respect to $t$ leads to
\begin{equation}
\label{eq:branching_rate}
\frac{d K_d}{dt} 
\sim n_d^{-B} \frac{d n_d}{dt} \sim p(n_d)\frac{d n_d}{dt}.
\end{equation}
In the last expression we assumed that the branching factor increases whenever a game arriving at the vertex generates a new branch.
The arrival of new games at the vertex at depth $d$ occurs at rate $dn_d/dt$, and the generation of new branches with probability $p(n_d)$.
Therefore, the approximate relation, $\nu \simeq B$, is obtained.
We should remark that our theoretical considerations correspond to a mean field approximation, in the more general case $\nu$ and $B$ might be different.
However, as we will show with our numerical simulations, the approximation, $\nu \simeq B$, is good enough when $B\simeq 0.85$ and it is robust under model variations.
The fact that the growth exponent $\nu$ is independent of $d$ is consistent with a self-similar growth process where at each vertex the same stochastic mechanism operates. Note that the innovation rate per vertex, $dK_d/dt$, is not constant as opposed to the constant growth rate in the standard formulation of the Yule-Simon preferential growth process~\cite{yule1925mathematical,simon1955class}.

\begin{figure}[!t]
\includegraphics*[width=\linewidth]{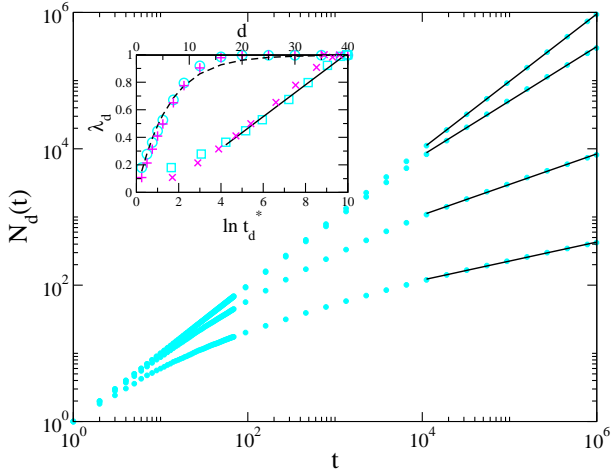}
\caption{\label{fig:4}
(Color online).
Simulation results for the models implementing the nested preferential growth mechanism.
Tree width, $N_d$, is plotted as a function of the number of games, $t$, and depths, $d=2,4,9,16$, for the model with linear preferential growth probability (cyan circles).
The data is logarithmically binned for $t > 100$.
The black solid lines are fitted results of the form $N_d = t_d^*(t/t_d^*) t^{\lambda_d}$.
{\em Inset:} 
The estimated values of $\lambda_d$ are plotted as a function of $d$ for the non-linear (linear) preferential growth probability using magenta plus signs (cyan circles), and as a function of the estimated values of $\ln(t_d^*)$ using magenta crossings (cyan squares).
The black dashed line corresponds to $\lambda_d = 1-B^d$ with $B=0.846\pm0.005$ ($R^2=0.98$), and the black solid line to $\lambda_d = a+(1/A) \ln t_d^*$ where $a=-0.15\pm0.02$ and $1/A=0.116\pm0.003$ ($R^2=0.996$).
Both are fits to the curves of the linear preferential growth case.
}
\end{figure}

Based on the theoretical considerations, we propose a generative mechanism consisting of nested preferential growth processes. 
We grow an ensemble of one hundred game trees each generated by one million games.
Then, we calculate $\lambda_d$ as a function of $d$ and $t_d^*$ from the average of $N_d(t)$, in order to compare the results with the empirical case (see fig.~\figu{fig:4}).
Each simulation starts from a tree with the root vertex only, and games are added one by one.
If the $(t+1)$-th incorporated game reaches a vertex $v$ with frequency $n_v(t)$, the vertex $v$ generates a new child vertex with probability $p(n_v(t))$. 
Otherwise, with probability $1-p(n_v(t))$, the game continues to one of the existing child vertices of $v$.
We use two types of preferential growth probabilities.
A non-linear preferential growth according to eq.~\equ{eq:pref_growth}, and a linear preferential growth given by $\pi(n|n') \propto n$.
More specifically, when there is not a branching event, a move from a vertex $v$ to one of its child vertices $u$ is performed with probability, $\pi(n_u|n_v) = q(n_u/n_v) / [\sum_{u'} q(n_{u'}/n_v)]$, for the non-linear case, and with probability, $\pi(n_u|n_v) = n_u/n_v$, for the linear case.
After the game traverse the tree, the vertex frequencies in the corresponding path are increased by one.
In the simulations, we choose the value of the parameter $\nu$ that provides the best prediction for $B = 0.85$ in order to reproduce the empirical case.
For the non-linear preferential growth probability, the best prediction of $B$ occurs at $\nu=0.95$, and in the linear case at $\nu=0.85$ (Inset in fig.~\figu{fig:4}, plus signs and circles, respectively).
As we previously pointed out, in both cases the approximation, $B \simeq \nu$, holds.
Moreover, the scaling behavior between the crossover point $t_d^*$ and $\lambda_d$ properly reproduce the empirical case (Inset in fig.~\figu{fig:4}, crosses and squares, respectively).
In the absence of preference, i.e. $\pi(n|n')$ independent of $n$, the simulation results deviates significantly from the empirical case (not shown).
Finally, the proposed mechanism turns out to be robust against variations of its details. For example, the replacement of eq.~\equ{eq:prob_branch} by fragmentation processes~\cite{blasius2009zipf} or the introduction of noisy selection of child vertices~\cite{zanette2005dynamics} leads to qualitatively the same results (not shown).


\section{Discussion}
The self-similar nature of the game tree and its generative mechanism implies a lack of typical scales of the innovation phenomena in chess. 
Equation~\equ{eq:prob_branch} indicates that there are no vertices with particularly large frequencies after which further innovation becomes impossible.
In other words, in chess there is no winning strategy, but there is always a possibility for innovative solutions to be introduced. 
Moreover, the observed preferential growth mechanism suggests that the exploitation also works in a self-similar way. 

Our findings in chess exhibit similarities to vocabulary growth in language evolution. 
The crossovers in the Heaps' law in eq.~\equ{eq:crossover} have a direct interpretation in vocabulary growth. 
According to Gerlach and Altmann~\cite{gerlach2013stochastic}, such crossover is rooted in the existence of two types of words, core words and non-core words, due to a separation of time scales in the language evolution process. 
This suggests the existence of core and non-core move sequences in the game tree.
However, notice that the dataset do not contain games from the beginning of chess-playing, but from the year 1998.
Therefore, the initial linear growth of $N_d(t)$ might be the consequence of non-realistic innovations due to random fluctuations for small values of $t$.
Nevertheless, our simulations shown in fig.~\figu{fig:4} do not implement a delayed beginning of the measurement process of the tree growth, and still exhibit a transition between two regimes predicting the right relationship between the crossover point $t_d^*$ and $\lambda_d$ (eq.~\ref{eq:crossover}).
Moreover, we simulated a delayed start of the measurement process by using the last $10^6$ games generated from a simulation with $10^7$ games.
We found no significant differences with the results in fig.~\figu{fig:4} (not shown).
Therefore, the crossover in the Heaps' law and the existence of core and non-core moves are intrinsic properties of the the tree evolution.
For the long time behavior, it has been found that the Heaps' exponent is larger in languages with a larger degree of inflection, where through declination and conjugation several words may be generated from root words~\cite{zanette2005dynamics}. 
This is consistent with our results if we make an analogy between the degree of inflection and a depth in the tree, because $\lambda_d$ grows with $d$. 
Thus, we provide further evidence about the origin of the non-universal Zipf's exponents~\cite{eliazar2011growth}.

\section{Conclusions}
In this work we studied how innovations are introduced into chess games by analyzing the evolution of the game tree. 
In our picture, move sequences are in one-to-one correspondence with the vertices in the tree.
The probability that a new innovation event occurs at a vertex decays as a power law of the frequency at which the vertex is reached.
Already known move sequences are played or exploited according to their frequencies, following a preferential growth process.
Our model is consistent with previous results on the static properties of the already explored game tree~\cite{blasius2009zipf}, and introduces important clues about its growing dynamics.
We found striking similarities between the evolution of the chess game tree and vocabulary growth. 
Based on our empirical findings, we proposed a generative mechanism that reproduces the observations and is robust with respect to variations of its details. 
All these findings provide insights into innovation phenomena in the context of decision making.

We acknowledge J. Saram\"aki for the careful reading of the manuscript. Financial support from the Academy of Finland, project No. 260427 (J.I.P.) and from Aalto University postdoctoral program (H.J.) is acknowledged. This work was partially supported by grants from CONICET, and SeCyT Universidad Nacional de C\'ordoba (Argentina).

\bibliographystyle{plain}

\end{document}